\documentclass[aps,twocolumn,secnumarabic,nobalancelastpage,amsmath,amssymb,
footinbib,superscriptaddress]{revtex4-1}

\usepackage{graphics}      % standard graphics specifications
\usepackage{graphicx}      % alternative graphics specifications
\usepackage{longtable}     % helps with long table options
\usepackage{url}           % for on-line citations
\usepackage{bm}            % special 'bold-math' package
\usepackage{physics}
\usepackage{xcolor}
\usepackage{float}
\usepackage[percent,grid,tics=10]{overpic}

\newcommand\numberthis{\addtocounter{equation}{1}\tag{\theequation}}
\newcommand\ddfrac[2]{\frac{\displaystyle #1}{\displaystyle #2}}

\begin{document}
\setlength{\parskip}{0pt}
\title{Ultrafast relaxation of symmetry-breaking photo-induced atomic forces}
\author         {Shane M. O'Mahony}
\affiliation    {Department of Physics, University College Cork, Cork T12K8AF, Ireland}
\affiliation    {Tyndall National Institute, Cork T12R5CP, Ireland}
\author         {Felipe Murphy-Armando}
\affiliation    {Tyndall National Institute, Cork T12R5CP, Ireland}
\author         {\'Eamonn D. Murray}
\affiliation     {Department of Physics and Department of Materials, Imperial College London, London SW7 2AZ, United Kingdom}
\author         {Jos\'e D. Querales-Flores}
\affiliation    {Tyndall National Institute, Cork T12R5CP, Ireland}
\author        {Ivana Savi\'c}
\affiliation    {Tyndall National Institute, Cork T12R5CP, Ireland}
\author         {Stephen Fahy}
\affiliation    {Department of Physics, University College Cork, Cork T12K8AF, Ireland}
\affiliation    {Tyndall National Institute, Cork T12R5CP, Ireland}
\email          {shane.m.omahony@umail.ucc.ie}
\date{\today}

\begin{abstract}
We present a first-principles method for the calculation of the temperature-dependent relaxation of symmetry-breaking atomic driving forces in photoexcited systems. We calculate the phonon-assisted decay of the photoexcited force on the low-symmetry $E_g$  mode following absorption of an ultrafast pulse in the prototypical group-V semimetals, Bi, Sb and As. The force decay lifetimes for Bi and Sb are of the order of $10$ fs and in good agreement with recent experiments, demonstrating that electron-phonon scattering is the dominant mechanism relaxing the symmetry-breaking forces. Calculations for a range of absorbed photon energies suggest that larger amplitude,  symmetry-breaking atomic motion may be induced by choosing a pump photon energy which maximises the product of the initial $E_g$ force and its lifetime.  We also find that the high-symmetry $A_{1g}$ force undergoes a partial decay to a non-zero constant on similar timescales, which has not yet been measured in experiments. We observe that the imaginary part of the electron self-energy, averaged over the photoexcited carrier distribution, provides a reasonable estimate for the decay rate of symmetry-breaking forces. 
\end{abstract}

\maketitle

%, including photocatalysis
%\cite{Stahler2008}, efficient renewable energy, laser annealing of materials,
%and, increasingly, in furthering our understanding of
%strongly correlated and charge-density-wave systems \cite{Schmitt2008}.

%%%%%%%%%%%%%%%%%%%%%%%%%%%%%%%%%%%%%%%%%%%%%%%%%%%%%%%%%%%%%%%%%%%%%%%%%%%%%

The generation and control of atomic forces in optically excited molecules and materials is important for a number of areas including photocatalysis \cite{Stahler2008}, laser annealing and the study of photo-assisted phase transitions~\cite{Wall572, Teitelbaum_2018}, with applications that include the development of efficient renewable energy~\cite{Zhongguo_2019} and phase-change memories~\cite{Salinga_2018, Kuramochi_2015}. 
The development of ultrafast optical spectroscopy has greatly advanced our understanding of electron and phonon dynamics in optically excited materials, with time
resolution on the tens-of-femtoseconds scale readily accessible~\cite{Shah1996}. More recently, time-resolved x-ray diffraction and time-resolved photoemission spectroscopy have
allowed the direct observation of atomic motion and electronic dynamics on timescales shorter than a picosecond following photoexcitation \cite{Fritz2007, Johnson2013, Trigo2013, Stolow2004}, providing insight into the physics of strongly correlated and charge density wave systems \cite{Schmitt2008,Mansart5603}. 

Photoexcitation with a laser pulse of duration much less than the fastest phonon period can be used to launch large amplitude coherent atomic motion in a variety of materials and molecules \cite{Fritz2007}. However, symmetry-breaking coherent atomic motion has been shown experimentally to have an amplitude orders of magnitude less than that of symmetry-preserving coherent atomic motion in a variety of materials \cite{Li2013, Misochko2015, Kamaraju2010, Huber2015}. Furthermore, the amplitudes of symmetry-breaking modes decrease strongly with increasing sample temperature, whereas high-symmetry mode amplitude is relatively insensitive to temperature, indicating very different aspects of the ultrafast dynamics affecting the two cases. If we understand the limiting mechanisms, it may be possible to suppress them and drive larger amplitude symmetry-breaking coherent atomic motion.

Bi and Sb are useful model systems for pump-probe reflectivity experiments due to their large
vibrational response to optical excitation \cite{Stevens2002, Garrett1996, Cheng1991, Crespo-Hernandez2004}.
High-symmetry coherent $A_{1g}$ phonons can be
generated through a mechanism termed displacive excitation
of coherent phonons (DECP) \cite{Zeiger1992}, related to the absorptive part of the Raman response \cite{Merlin1997}.
When the pump pulse is polarised perpendicular to the 3-fold rotational axis of the crystal, the symmetry-breaking $E_g$ mode has also been detected ~\cite{Johnson2013, Li2013}, but with a much lower and strongly temperature-dependent amplitude.  

%We don't use electron-phonon rate eqns to compute initial distribution. 
% Don't need to 
%We combined DFPT with elphon rate eqns to calculate the evolution of fs timescales of a photoexcited electronic dist gen by optical absorption and calculate the resulting
%atomic forces etc ....

In this work, we provide for the first time a quantitative understanding of how incoherent electron-phonon scattering limits the generation of symmetry-breaking coherent atomic motion. We combine density functional perturbation theory (DFPT) \cite{Baroni2001a}, and electron-phonon scattering rate equations \cite{Madelung1978} to calculate the evolution on fs timescales of a photoexcited electronic distribution generated by optical absorption and compute the resulting time-dependent atomic forces in the group-V semimetals, Bi, Sb and As.
We find that electron-phonon scattering dominates in determining the lifetime of the $E_g$ driving force in photoexcited Bi and Sb, with calculated lifetimes in good agreement with recent experiments \cite{Li2013}, and we predict similar behavior in As. 
We calculate the dependence of the initial atomic driving forces and their lifetimes on the photon energy of the pump pulse and suggest how variation of the incident photon energy may be used to maximise the impact on low-symmetry atomic motion.

Our method goes beyond standard time-dependent density functional theory (TDDFT) \cite{Runge1984} approaches by explicitly considering the coupling of the excited electron-hole plasma to the continuum of  thermal phonon modes throughout the Brillouin zone and can be used to compute the lifetime of symmetry-breaking photo-induced atomic forces on ultrafast timescales in a variety of materials. 

At room temperature, the $E_g$ mode in bismuth (antimony) was observed to have an amplitude $\sim 10$ ($30$) times smaller than the high-symmetry $A_{1g}$ mode  \cite{Johnson2013, Li2013}.  A density functional theory study has shown that the initial photo-induced driving force on the symmetry preserving ($A_{1g}$) and symmetry-breaking ($E_g$) coherent modes of bismuth are comparable \cite{Murray2015}. Therefore, the highly reduced amplitude of the $E_g$ mode indicates that the $E_g$ driving force is extremely short lived. Recent experimental work utilised a combination of optical pump-optical probe and continuous-wave (cw) Raman scattering to indirectly determine the lifetime of the $E_g$ driving force in Bi and Sb as a function of temperature. The $E_g$ force lifetime in bismuth (antimony) was found to vary from $13 \pm 4$ fs ($17 \pm 2$ fs) at $10$ K to $2 \pm 0.4$ fs ($5.5 \pm 0.5$ fs) at room temperature \cite{Li2013}. It was suggested that the rapid, temperature-dependent relaxation of this force was due to the initial low-symmetry excited electron-hole plasma rapidly regaining full symmetry via electron-phonon scattering. A similar conclusion was reached in other experimental work \cite{Johnson2013}, where the $E_g$ driving force in bismuth was shown to have a decay time of $\sim 4$ fs at room temperature. A study of the coherent modes in topological insulator $\text{Bi}_2\text{Te}_3$ showed similar behaviour of the symmetry-breaking $E_g$ modes, whose relatively small amplitude compared with the fully symmetric $A_{1g}$ modes was attributed to short-lived photoexcited electronic states with lifetimes $\sim 10$ fs \cite{Misochko2015}. A recent calculation \cite{ISI:000357060600023} found the timescale for equilibration of $L$ valley occupations in photoexcited silicon via electron-phonon scattering to be ~180 fs, but did not consider the consequences for the generation of symmetry-breaking coherent atomic motion.

\begin{figure}[h]
\includegraphics[width=8cm]{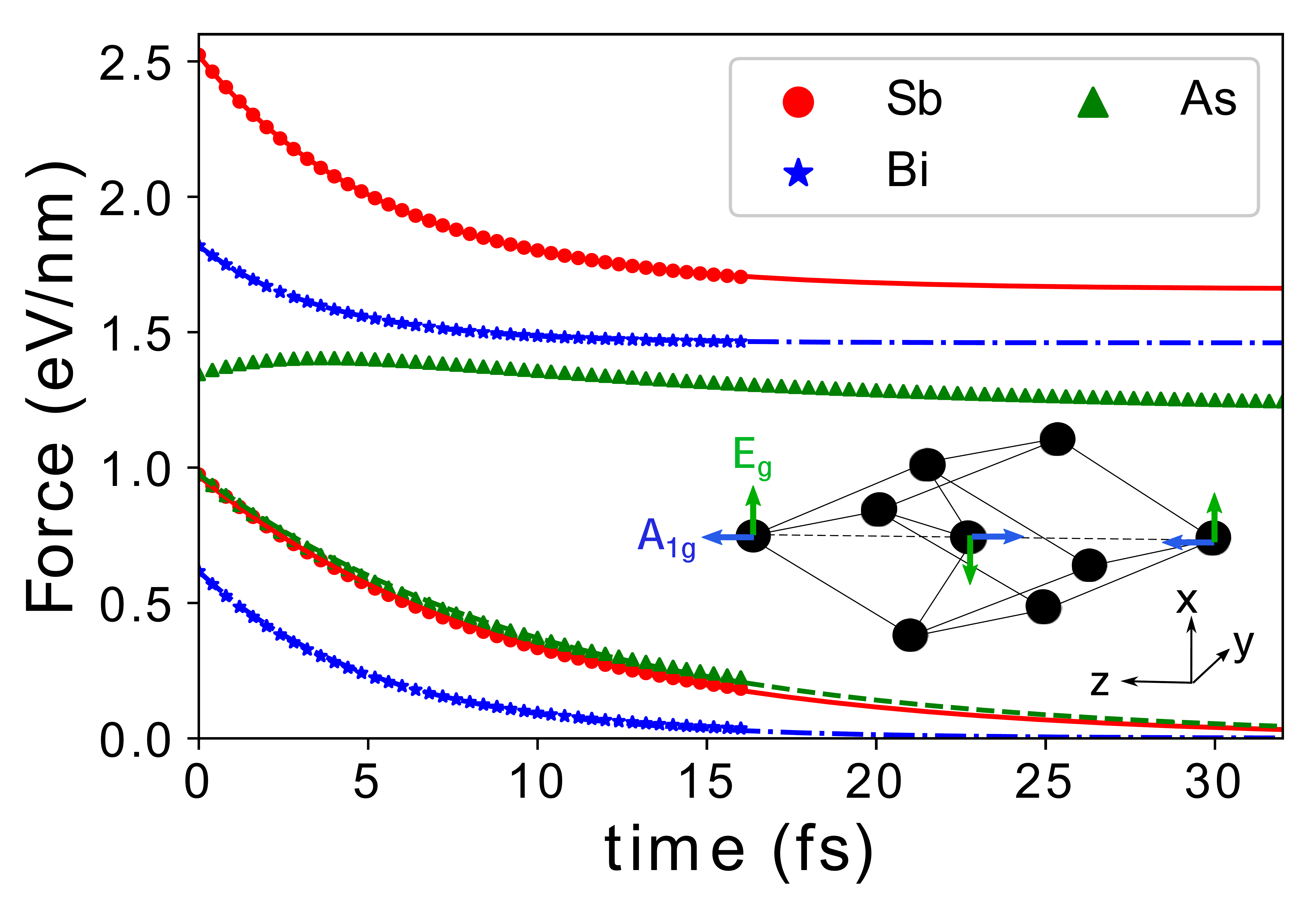}
\caption{The $E_g$ and $A_{1g}$ driving forces in Bi, Sb and As as functions of time-delay, following the absorption of 0.1 photons of energy $1.5$ eV per unit cell. The three upper plots show the $A_{1g}$ force, the three lower show the $E_g$ force. The solid lines are fittings to the explicitly calculated data points. $F_{E_g}$ is fit to a decaying exponential $F_{E_g}(t=0) \exp{-t/\tau_{E_g}}$ and $F_{A_{1g}}$ is fit to a decaying exponential plus a constant term:  $F_{A_{1g}}^1 \exp{-t/\tau_{A_{1g}}} + F_{A_{1g}}^2$ as discussed in the main text. Inset: The unit cell of the group-V semimetals. The green (blue) arrows indicate atomic motion corresponding to the $E_g$ ($A_{1g}$) modes.}
\label{Eg_A1g_force_v_time}
\end{figure}

The group-V semimetals crystallise in the A$7$ rhombohedral structure, with $2$ atoms per unit cell. One atom is at the origin and the other displaced a distance $zc$ along the trigonal axis ($c$-axis), which is represented by a dashed line in the inset of Fig. $\ref{Eg_A1g_force_v_time}$. The internal atomic displacement parameter, $z$,  is highly sensitive to excitation of electrons to the conduction bands. This alters the equilibrium value of $z$ and generates oscillations of the symmetry-preserving $A_{1g}$ mode.  In contrast, the symmetry-breaking $E_g$ mode involves motion of the atoms perpendicular to the $c$-axis and is thus not driven by the conventional DECP mechanism, which assumes occupations of excited electron states that preserve crystal symmetry. The $E_g$ mode is driven by unbalanced occupation of symmetry-equivalent regions of the Brillouin zone following photoexcitation by a pump polarised perpendicular to the $3$-fold axis of the crystal~\cite{Li2013,Murray2015}.  
%\color{blue} The amplitude of the $E_g$ mode in bismuth (antimony) has been experimentally determined to be $\sim 10$ ($30$) times smaller than that of the $A_{1g}$ mode \cite{Li2013}. The forces driving the $E_g$ and $A_{1g}$ modes have been shown to be comparable \cite{Murray2015}, suggesting that the discrepancy in the amplitudes is due to the short lifetime of the driving force on the $E_g$ mode. \color{black}

We compute electron states $| n{\bf k} \rangle$ with energy $\epsilon_{n{\bf k}}$ for band $n$ at momenta ${\bf k}$, and phonon normal modes ${\bf e} ^\lambda({\bf q})$ with frequency $\omega_{\lambda {\bf q}}$ at momenta ${\bf q}$ on a uniform grid in the Brillouin zone and find the electron-phonon matrix elements , $g_{\mathbf{k} n m}^{\lambda \mathbf{q}}$ as defined in Ref.~\cite{Giustino2017},  on the same grid using DFPT \cite{Baroni2001a}. These quantities are then interpolated to a finer grid using maximally localised Wannier functions \cite{Marzari2012}. We generate the initial photoexcited distribution in the same manner as Ref.~\cite{Murray2015}. The excited electronic occupations are then evolved in time using electron-phonon rate equations \cite{Madelung1978}:
%%%%%%%%%%%%%%%%%%%%%%%%%%%%%%%%%%%%%%%%%%%%%%%%%%%%%%%%%%%%%%%
\begin{align}
   &\frac{\partial f_{n \mathbf{k}}}{\partial t} = \sum_{m, \mathbf{q},\lambda, \xi} \left[ R^{\xi}_{\lambda} \left( m \mathbf{k} + \mathbf{q}, n \mathbf{k} \right) - R^{\xi}_{\lambda} \left( n \mathbf{k}, m \mathbf{k} + \mathbf{q} \right) \right] \\
   &\frac{\partial n_{\mathbf{q} \lambda}}{\partial t} = \sum_{\mathbf{k}, n, m} \left[ R^{+}_{\lambda} \left( m \mathbf{k} + \mathbf{q}, n \mathbf{k} \right) - R^{-}_{\lambda} \left( n \mathbf{k}, m \mathbf{k} + \mathbf{q} \right) \right], 
\end{align}
%%%%%%%%%%%%%%%%%%%%%%%%%%%%%%%%%%%%%%%%%%%%%%%%%%%%%%%%%%%%%%%
where $\xi =$ phonon emission $(+)$ or phonon absorption $(-)$, $f_{n \mathbf{k}}$ is the occupation of electronic state $\ket{n \mathbf{k}}$, $n_{\mathbf{q}, \lambda}$ are the phonon occupations for branch $\lambda$,  and $R^{\xi}_{\lambda}$ are the electron-phonon scattering rates defined by Fermi's golden rule: 
%%%%%%%%%%%%%%%%%%%%%%%%%%%%%%%%%%%%%%%%%%%%%%%%%%%%%%%%%%%%%%%
\begin{align*}
    &R^{\pm}_{\lambda} \left(n \mathbf{k}, m \mathbf{k} + \mathbf{q} \right) = \frac{1}{N}  \frac{1}{\omega_{\lambda \mathbf{q}}} \left|g_{\mathbf{k}  n m}^{\lambda \mathbf{q}} \right|^2 f_{n\mathbf{k}} \quad \times  \\
     &\left( 1 - f_{m \mathbf{k} + \mathbf{q}} \right) \left(n_{\lambda \mathbf{q}} + \frac{1}{2} \pm \frac{1}{2} \right)  \delta(\varepsilon_{m \mathbf{k} + \mathbf{q}} - \varepsilon_{n \mathbf{k}} \pm \omega_{\lambda \mathbf{q}})  \numberthis  \label{rate-equation}, 
\end{align*}
%%%%%%%%%%%%%%%%%%%%%%%%%%%%%%%%%%%%%%%%%%%%%%%%%%%%%%%%%%%%%%%
where $N$ is the number of wave vectors $\bf k$ (or {$\bf q$}) in the uniform Brillouin zone grid, $\omega_{\mathbf{q} \lambda}$ are the phonon frequencies and $\delta(\varepsilon_{m \mathbf{k} + \mathbf{q}} - \varepsilon_{n \mathbf{k}} \pm \omega_{\lambda \mathbf{q}})$ are the energy conserving delta functions for emission and absorption of a phonon. Finite lifetimes give the electronic states a Lorentzian line shape in energy and the energy conservation delta-function broadens to a Lorentzian whose width is the sum of the linewidths of the initial and final state in the scattering process \cite{Marini_2013}: 
%%%%%%%%%%%%%%%%%%%%%%%%%%%%%%%%%%%%%%%%%%%%%%%%%%%%%%%%%%%%%%%
\begin{equation}
 P^{\pm} = \frac{\Im{\Sigma_{m \mathbf{k} + \mathbf{q}} + \Sigma_{n \mathbf{k}}}}{[\Delta{\varepsilon_{\mathbf{k}, \mathbf{k} + \mathbf{q}}^{nm}} \pm \omega_{ \lambda \mathbf{q}}]^2 + [\Im{\Sigma_{m \mathbf{k} + \mathbf{q}} + \Sigma_{n \mathbf{k}}}]^2}, 
\end{equation}
%%%%%%%%%%%%%%%%%%%%%%%%%%%%%%%%%%%%%%%%%%%%%%%%%%%%%%%%%%%%%%%
where $\Delta{\varepsilon_{\mathbf{k}, \mathbf{k} + \mathbf{q}}^{nm}} = \varepsilon_{m \mathbf{k} + \mathbf{q}} - \varepsilon_{n \mathbf{k}} $ and  $\Im{\Sigma_{n \mathbf{k}}}$ is the imaginary part of the electron self-energy for state $\ket{n \mathbf{k}}$ . It is temperature-dependent and related to the equilibrium lifetime of the state via $1/\tau_{n \mathbf{k}} (T)= 2 \Im{\Sigma_{n \mathbf{k}}(T)}/\hbar$. \cite{Giustino2017} \footnote{In practice, we calculate $\Im{\Sigma_{n \mathbf{k}}}$ by replacing the energy conserving delta function with a Gaussian. However, the calculated $E_g$ force lifetimes are insensitive to the width of this Gaussian.}. This brings our rate equations into agreement with the completed-collisions limit of the Kadanoff-Baym equations  \cite{ISI:000357060600023}. For reasons of numerical efficiency, we replace these Lorentzians with Gaussians of the same width.

The atomic force $\bm{F_{\alpha }}$ on atom $\alpha$ in the unit cell is computed at each time step using the diagonal part of the electron-phonon matrix:
%%%%%%%%%%%%%%%%%%%%%%%%%%%%%%%%%%%%%%%%%%%%%%%%%%%%%%%%%%%%%%%
\begin{equation}
    \bm{F_{\alpha }} = - \frac{1}{N} \sum_{n,\mathbf{k}} \Delta f_{n \mathbf{k}}  \mel{ n \mathbf{k}} {\nabla_{\bm{\tau_\alpha}} \hat{H}}{n \mathbf{k}}, 
    \label{force equation}
\end{equation}
%%%%%%%%%%%%%%%%%%%%%%%%%%%%%%%%%%%%%%%%%%%%%%%%%%%%%%%%%%%%%%%
 where $\Delta f_{n \mathbf{k}} = f_{n \mathbf{k}} - f_{n \mathbf{k}}^{0}$ is the change in occupation of state $\ket{n \mathbf{k}}$ from its equilibrium value and $\bm{\tau_\alpha}$ is the displacement of atom $\alpha$ from equilibrium. \footnote{We have also computed the atomic forces self consistently within the framework of CDFT \cite{Tangney1999} using the ABINIT package \cite{Gonze2009}. We find that the $E_g$ force computed both ways agree very well, so we use the approach in Eq. $\eqref{force equation}$ as it is more efficient. The $A_{1g}$ force depends more delicately on the exact values of the equilibrium electronic occupations, $f_{n \mathbf{k}}^{0}$, so is more accurately computed using CDFT.} 

The time evolution of both the $E_g$ and the $A_{1g}$ driving forces are shown in Fig.~$\ref{Eg_A1g_force_v_time}$,  demonstrating that the $E_g$ force exponentially decays to zero, as expected, while the $A_{1g}$ force undergoes a more complex time evolution. In Bi and Sb, the $A_{1g}$ force undergoes a partial decay from its initial value to a non-zero constant. In As, the $A_{1g}$ force increases slightly before decaying to a non-zero constant. The final values of the $A_{1g}$ forces in Bi and Sb are obtained by fitting the calculated time-dependent values to a decaying exponential plus a constant term, as explained in the caption of Fig.~$\ref{Eg_A1g_force_v_time}$. In As, we extract the final $A_{1g}$ force by fitting the same function to the tail of the calculated values. In Table~\ref{table:forces}, we see that the final $A_{1g}$ forces in all three materials are slightly higher than that which would be obtained in constrained DFT (CDFT) by assuming a hot thermal distribution of electrons and holes, with different chemical potentials~\cite{Tangney1999}. The $A_{1g}$ force will eventually relax to $0$ when the excited electronic occupations return to equilibrium i.e.  $\Delta f_{n \mathbf{k}} = 0$, as indicated by Eq. $\eqref{force equation}$. However, this process occurs on much longer timescales ($> 10$ ps) \cite{Sheu2013} and is beyond the scope of this work. 

\begin{table}[ht]
\caption{Comparison of initial ($F_{A1g}^{i}$) and final $A_{1g}$ force ($F_{A1g}^{f}$) with those obtained in a two-chemical potential CDFT calculation ($F_{A1g}^{2 \mu}$)~\cite{Tangney1999}. The forces are computed assuming an absorbed fluence of 0.1 photons of energy $1.5$ eV per unit cell.}
\centering
\begin{tabular}{c c c c}
\hline\hline
Material & $F_{A1g}^{i}$ (eV/nm) & $F_{A1g}^{f}$ (eV/nm) & $F_{A1g}^{2 \mu}$ (eV/nm) \\ [0.5ex] % inserts table %heading
\hline
Bismuth & 1.82 & 1.46 & 1.34 \\
Antimony & 2.52 & 1.66 & 1.38 \\
Arsenic & 1.35 & 1.21 & 1.13 \\ [1ex]
\hline
\end{tabular}
\label{table:forces}
\end{table}

%to fully understand it we would need to accurately account for the diffusion of the excited carriers away from the region defined by the product of the optical absorption depth and the pump spot size. In addition, once the electrons cool to the conduction band extrema,  we would need to treat the electron-hole recombination between the $\mathbf{T}$ hole band and the $\mathbf{L}$ electron band, a regime in which the phase space for electron-phonon scattering is extremely restricted. A realistic treatment of this regime is beyond the scope of this work. 

%\begin{figure*}
%\begin{overpic}[tics=5,width=15cm]{./figures/output.pdf} 
%\put(25,27){\color{blue} (a)}
%\end{overpic}
%\caption{Lifetime of driving force on $E_g$ mode as a function of lattice temperature. The red line is the theoretical result, the black points are the experimental measurements [Li et.al 2013]} 
%\end{figure*}

%\begin{figure*}[t!]
%\includegraphics[width=15cm]{./figures/output.pdf}
%\caption{Lifetime of driving force on $E_g$ mode as a function of lattice temperature. The red line is the theoretical result, the black points are the experimental measurements [Li et.al 2013]} 
%\end{figure*}

\begin{figure}[t]
\includegraphics[width=8cm]{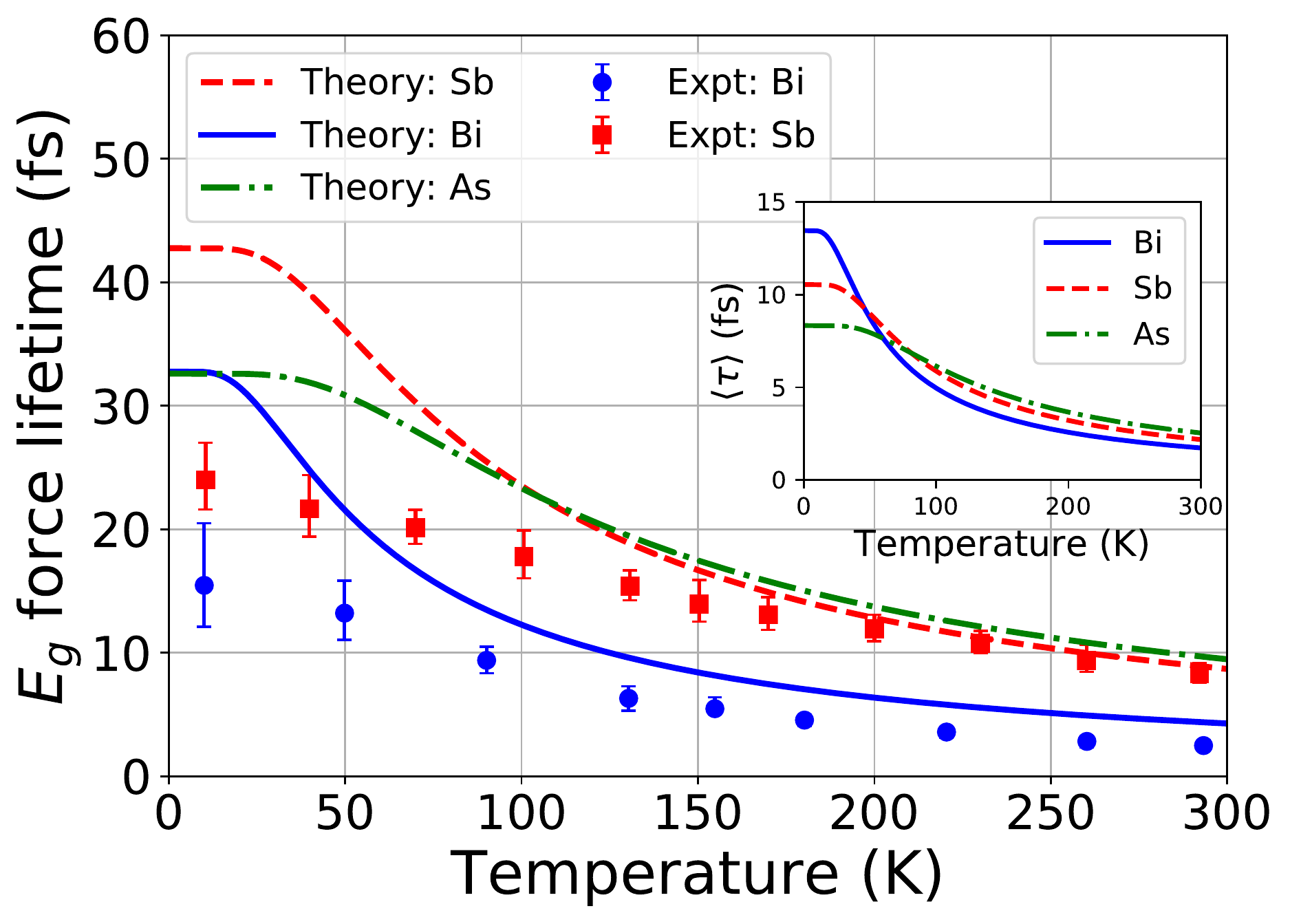}
\caption{Lifetime of driving force on $E_g$ mode as a function of lattice temperature for Bi, Sb and As for a pump photon energy of $1.5$ eV. The curves are theoretical results, the points are the experimentally inferred values \cite{Li2013} re-derived taking into account the partial decay of the $A_{1g}$ force. Inset: the average lifetime of states within the excited electron-hole plasma, as defined in Eq. $\eqref{average_state_lifetime}$. Both $\langle \tau \rangle$ and $\tau_{E_g}$ were computed for 16 temperatures in the interval $[0.1,300]$ K and fitted with the function $f(T) = f(0)/[1 + 2n_B (T,\Omega_0)]$, where $n_B(T, \Omega_0)$ is the Bose-Einstein occupation number for a mode frequency $\Omega_0$ at temperature $T$ and $\Omega_0$ is a fitting parameter: $\hbar \Omega_0(\text{Bi}) \approx 6.8$ meV, $\hbar \Omega_0(\text{Sb}) \approx 10.7$ meV and $\hbar \Omega_0(\text{As}) \approx 15.5$ meV. }
\label{Eg_force_decay}
\end{figure} 

Li et. al.  \cite{Li2013} determined the $E_g$ force lifetime indirectly, by comparing the ratio of the $E_g$ to $A_{1g}$ mode amplitude in an optical pump-optical probe experiment with cross-sections obtained in a cw Raman scattering experiment. The experimental $E_g$ force lifetimes in Ref.~\cite{Li2013} were derived, assuming that the $A_{1g}$ force does not change over the duration of the pump-pulse ($\sim 70$ fs). However, our calculations show a decay of the $A_{1g}$ force from $F_0 \rightarrow sF_0$ in much less than $70$ fs for Bi ($s \sim 0.80$) and Sb ($s \sim 0.65$), as shown in Fig. $\ref{Eg_A1g_force_v_time}$. We adjust the experimental analysis in Ref.~\cite{Li2013} to account for this partial decay. See appendix D for the full details.  This allows us to make a quantitative comparison between our calculated $E_g$ force lifetimes and the experimental ones.  

In Fig. $\ref{Eg_force_decay}$, the calculated and experimental $E_g$ force lifetime are shown as functions of temperature for Bi and Sb~\footnote{Electron energies, phonon frequencies and electron-phonon matrix elements were computed on a $6 \times 6 \times 6$ grid using Quantum Espresso \cite{ESPRESSO2009} and interpolated to a $14 \times 14 \times 14$ grid using the EPW code. \cite{Ponce2016}}, and only the calculated values for As, where no experimental measurements are available. The pump pulse photon energy in the calculations is $1.5$ eV, as in the experiment in Ref.~\cite{Li2013}. 

The agreement between theory and experiment is very good. In particular, the calculated $E_g$ force relaxation rate in antimony differs from experiment by a constant scattering rate of $\Gamma \sim 12.5 \hspace{1mm} \text{ps}^{-1}$, consistent with a temperature-independent scattering mechanism due to static imperfections in the sample, such as impurities or grain boundaries. The calculated relaxation rate in bismuth differs from experiment by a roughly uniform factor of $\sim 1.5$. This small discrepancy could be due to some additional scattering mechanism not considered here, such as electron-electron scattering. However, given the challenging nature of the experiment, the agreement is still excellent and confirms that electron-phonon scattering is the dominant relaxation mechanism for the $E_g$ driving force in both materials. 

The inset of Fig. $\ref{Eg_force_decay}$ shows the average lifetime of states within the excited electron-hole plasma due to electron-phonon coupling, which we define: 
%%%%%%%%%%%%%%%%%%%%%%%%%%%%%%%%%%%%%%%%%%%%%%%%%%%%%%%%%%%%%%%
\begin{equation}
   \frac{1}{\langle \tau(T) \rangle} = \ddfrac{\sum_{n \mathbf{k} \in \text{vb}} \gamma_{n \mathbf{k}}(T) (1 - f_{n \mathbf{k}})}{\sum_{n \mathbf{k} \in \text{vb}} (1 - f_{n \mathbf{k}})}  +  \ddfrac{\sum_{n \mathbf{k} \in \text{cb}} \gamma_{n \mathbf{k}}(T) f_{n \mathbf{k}}}{\sum_{n \mathbf{k} \in \text{cb}} f_{n \mathbf{k}}},
    \label{average_state_lifetime}
\end{equation}
%%%%%%%%%%%%%%%%%%%%%%%%%%%%%%%%%%%%%%%%%%%%%%%%%%%%%%%%%%%%%%%
where $\gamma_{n \mathbf{k}}$ are the equiibrium inverse relaxation times of the electronic states $\ket{n \mathbf{k}}$ \cite{Grimvall} and $f_{n \mathbf{k}}$ are the initial photoexcited electronic occupations following absorption of $1.5$ eV photons. We see that the average lifetime of the electron-hole plasma is similar to, but less than the $E_g$ force lifetime in all three materials, since not all electron-phonon scattering events degrade the $E_g$ force, but all relax the states within the electron-hole plasma. We further note that the temperature dependence of the $E_g$ force lifetime is very similar to that of $\langle \tau \rangle$. 

Thus, if we know the lifetime of the $E_g$ force at low-temperature, we can make a good estimate of $\tau_{E_g}(T)$ by computing $\langle \tau (T)\rangle$, which is computationally much less demanding. In more structurally complex materials, where a full simulation of the force decay might be very difficult, $\langle \tau \rangle$ should provide a reasonable approximation of the lifetime of symmetry-breaking atomic driving forces. 
 
 The similarity between $\langle \tau \rangle$ and the lifetime of the $E_g$ force underlines the importance of the coupling between the excited electron-hole plasma and the continuum of thermal vibrations in determining the behaviour of symmetry-breaking atomic driving forces. This is in contrast to the $A_{1g}$ driving force, which decays to a value similar to the force we would obtain by assuming two separate thermal distributions for the photoexcited electrons and holes, on timescales less than $100$ fs (as shown in Fig. $\ref{Eg_A1g_force_v_time}$ and Table \ref{table:forces}). This underlines that the dynamics affecting symmetry-breaking forces are quite different to those determining symmetry-preserving forces. 

\begin{figure}[h]
\includegraphics[width=8cm]{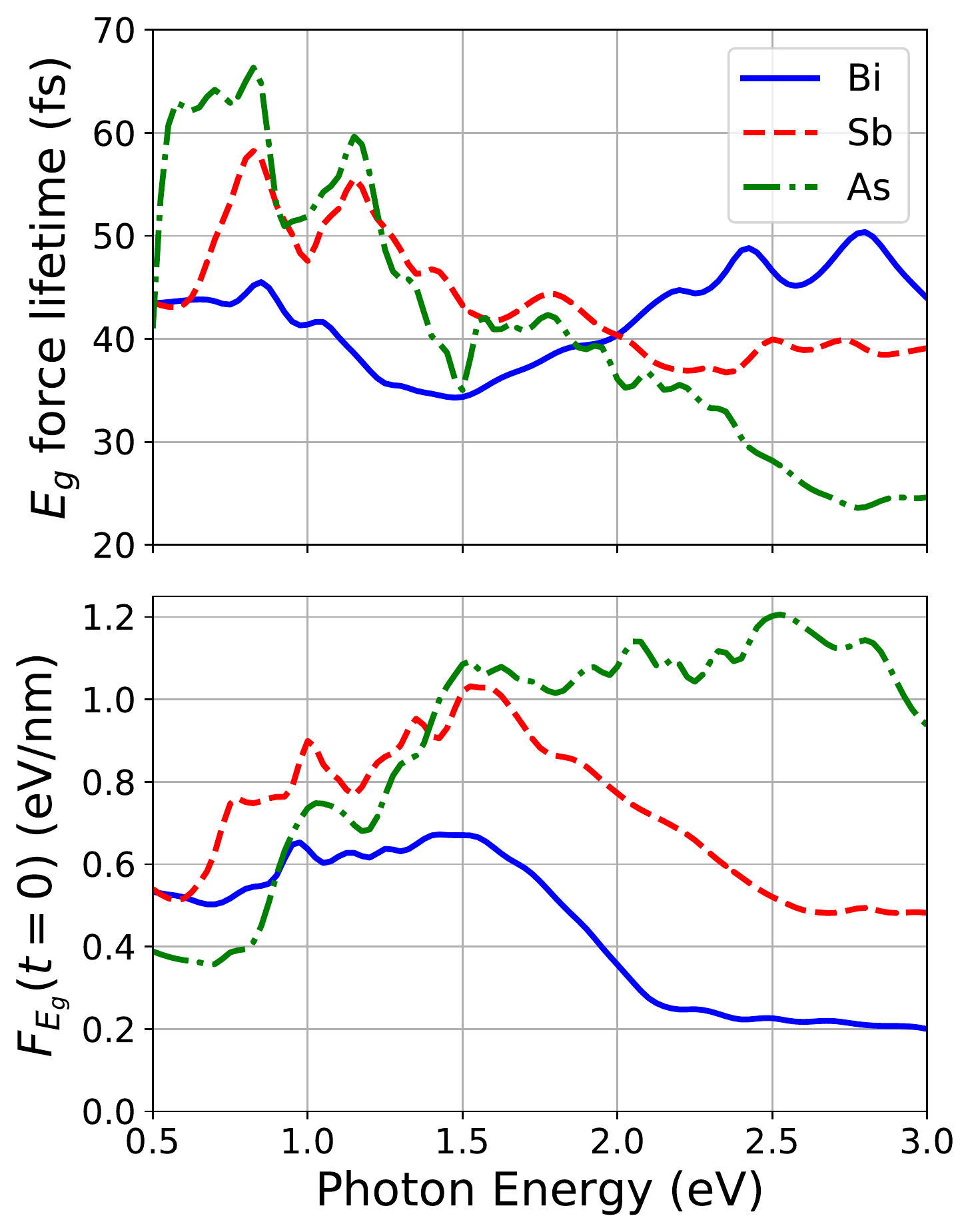}
\caption{Upper: Low temperature (0.1 K) driving force lifetime of the $E_g$ photoexcited force in Bi, Sb and As, as functions of the pump photon energy. Lower: Initial $E_g$ driving force as functions of pump photon energy, assuming $0.1$ photons absorbed per unit cell. }
\label{photon_energy}
\end{figure}

In Fig. $\ref{photon_energy}$, we show that both the initial $E_g$ force and the $E_g$ force lifetime vary substantially with the pump photon energy in all three materials \footnote{These calculations were performed on a $16 \times 16 \times 16$ grid}. The amplitude of the $E_g$ mode is proportional to the product $\tau_{E_g}F_{E_g}(t=0)$, for a given number of photons absorbed per unit cell~\footnote{A more exact expression would be $A_{E_g} \propto F_{E_g}(t=0)/\sqrt{1 + \frac{1}{\tau_{E_g}^2 \Omega_{E_g}^2}}$, which reduces to $A_{E_g} \propto \tau_{E_g}F_{E_g}(t=0)$ in the limit where $1/\tau_{E_g}^2 \Omega_{E_g}^2 \gg 1 $}. Within the energy range considered, this indicates that bismuth should be pumped with photons of energy $\sim 0.9$ eV, antimony with photons of energy in the interval $[1.0,1.5]$ eV and arsenic with photons of energy in the interval $[1.0,2.5]$ eV to maximise the $E_g$ mode amplitude. In other materials, choosing a photon energy which maximises $\tau_{E_g}F_{E_g}(t=0)$ would enable us to increase the amplitude of symmetry-breaking coherent modes, which could permit investigation into the possibility of inducing structural phase transitions which lower crystal symmetry.  \cite{Johnson2013}

In conclusion, we have presented a first principles method for calculating the generation and relaxation of low-symmetry photo-induced forces, which goes beyond conventional TDDFT approaches by explicitly considering coupling between the excited electron-hole plasma and the continuum of thermal vibrations, enabling us to accurately describe the ultrafast excitation and relaxation of the symmetry-breaking $E_g$ driving force in Bi, Sb and As. We have defined a non-equilibrium average lifetime of states within the electron-hole plasma and shown that it provides a reasonable estimate for the $E_g$ force lifetime and has the same temperature dependence as the $E_g$ force lifetime in all three materials, making it a computationally useful diagnostic for the lifetime of low-symmetry photo-induced forces in more structurally complex materials. We have demonstrated that the lifetimes of the $E_g$ forces in Bi, Sb and As vary substantially with the photon energy of the pump pulse, and suggest that similar effects would occur in other materials, providing a path to generating larger amplitude symmetry-breaking atomic motion by suitable choice of pump photon energy.  

This work was supported financially by Science Foundation Ireland award 12/IA/1601 and the Irish Research Council award GOIPG/2015/2784.

%%%%%%%%%%%%%%%%%%%%%%%%%%%%%%%%%%%%%%%%%%%%%%%%%%%%%%%%%%%%%%%%%%
\appendix
\section{Wannier interpolation of electron-phonon matrix elements}
\begin{figure}[h]
\includegraphics[width=8cm]{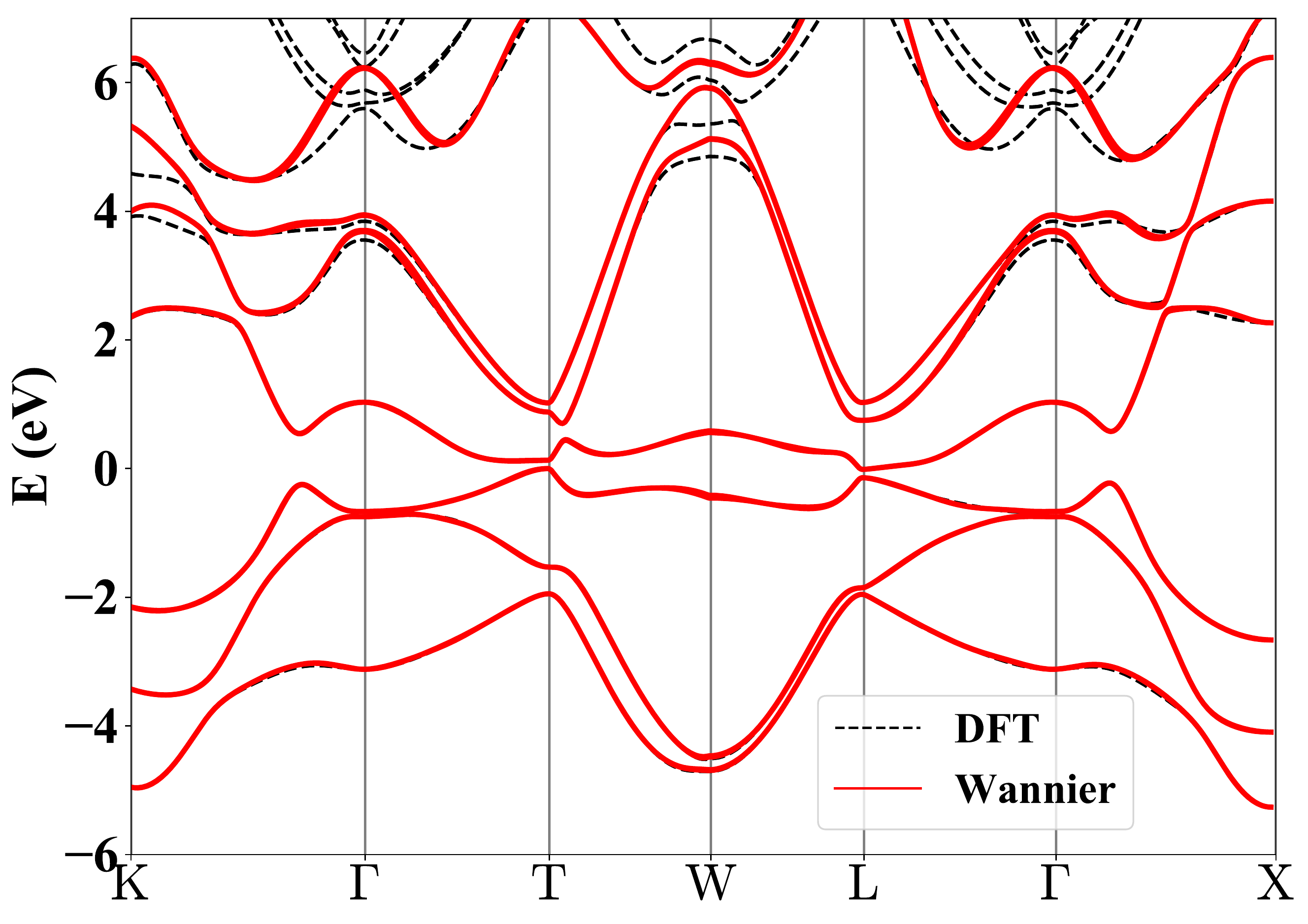}
\caption{Comparison of Bi DFT bands with those obtained by Wannier interpolation. 14 Wannier orbitals were used to interpolate the bandstructure from a coarse $6 \times 6 \times 6$ grid.}
\label{Bi_Wannier_bands}
\end{figure}

\begin{figure}[h]
\includegraphics[width=8cm]{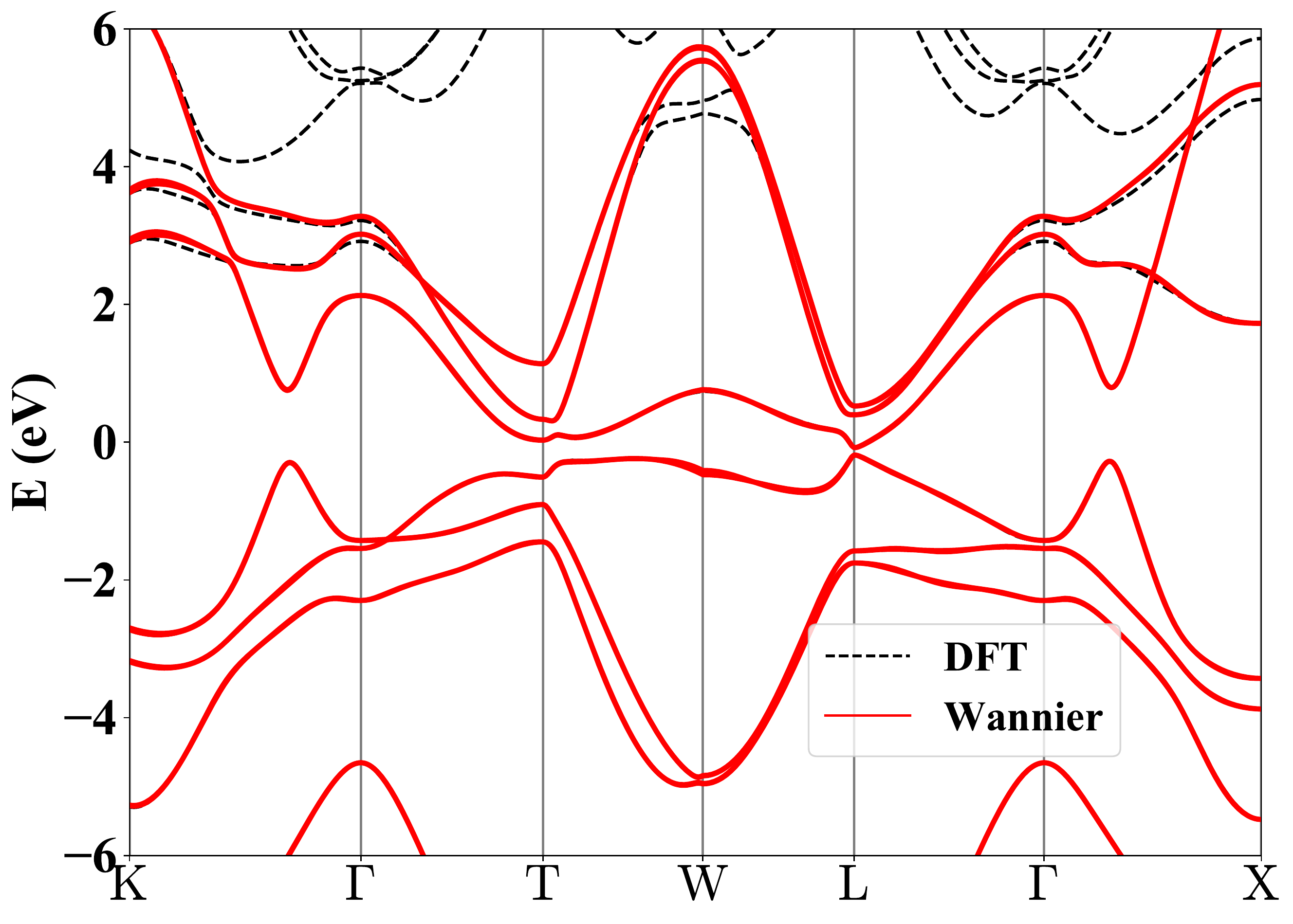}
\caption{Comparison of Sb DFT bands with those obtained by Wannier interpolation. 16 Wannier orbitals were used to interpolate the bandstructure from a coarse $6 \times 6 \times 6$ grid.}
\label{Sb_Wannier_bands}
\end{figure}

\begin{figure}[h]
\includegraphics[width=8cm]{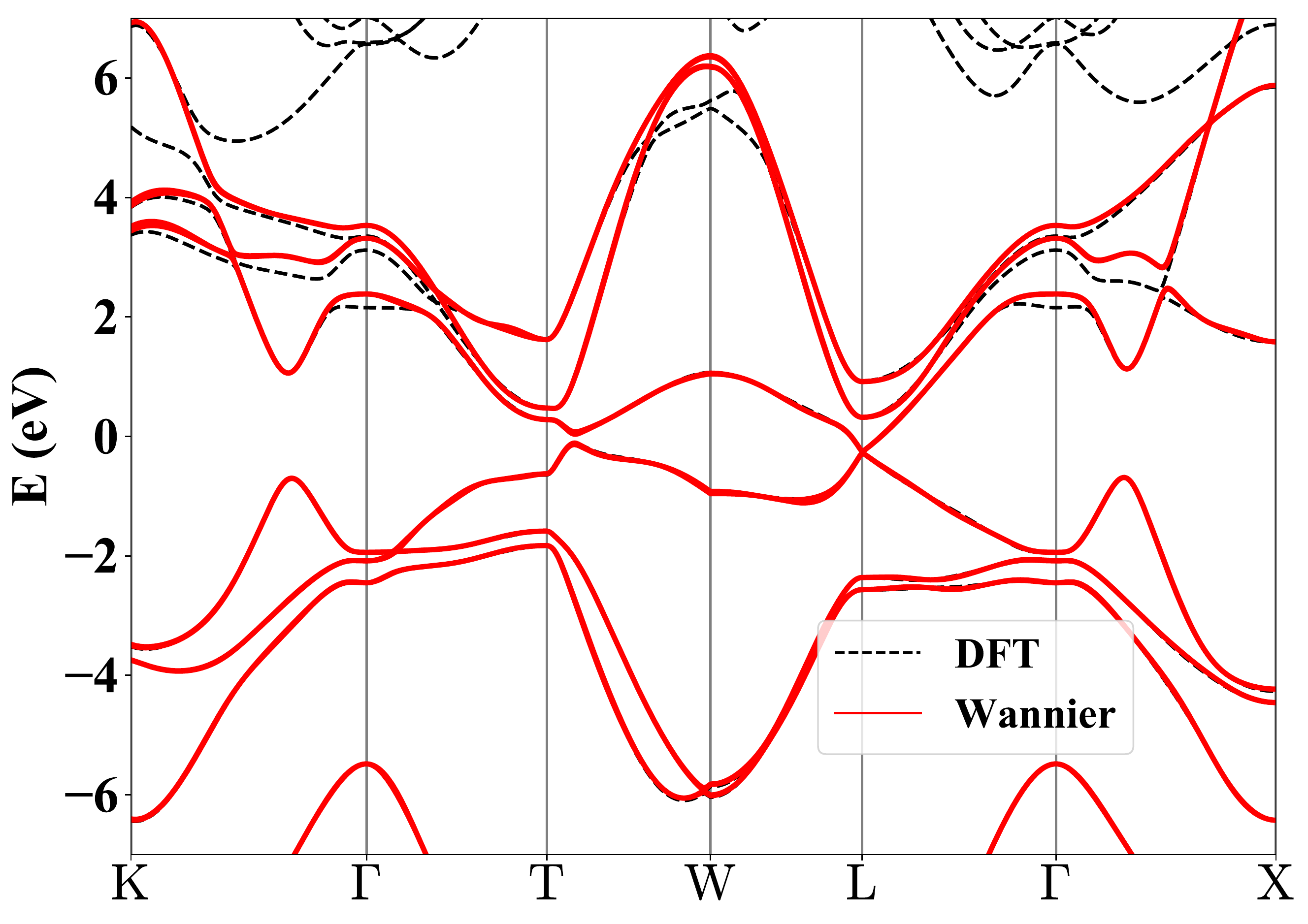}
\caption{Comparison of As DFT bands with those obtained by Wannier interpolation. 16 Wannier orbitals were used to interpolate the bandstructure from a coarse $6 \times 6 \times 6$ grid.}
\label{As_Wannier_bands}
\end{figure}

The electronic bandstructure, phonon dispersion and electron-phonon coupling matrix elements were calculated on a uniform $6 \times 6 \times 6$ Brillouin zone grid within the framework of density functional perturbation theory. We used a $20$ hartree plane wave energy cutoff and the local density approximation to exchange and correlation. Norm-conserving pseudopotentials including spin orbit coupling were used for all $3$ materials. These quantities were then interpolated to finer grids using maximally localised Wannier functions (MLWF) as implemented in the EPW code~\cite{Ponce2016}. The interpolation of the electronic bandstructure of Bi, Sb and As are shown in Figs.~$\ref{Bi_Wannier_bands}$, $\ref{Sb_Wannier_bands}$ and $\ref{As_Wannier_bands}$ respectively. 
%%%%%%%%%%%%%%%%%%%%%%%%%%%%%%%%%%%%%%%%%%%%%%%%%%%%%%%%%%%%%%%%%%%%%%%%
%%%%%%%%%%%%%%%%%%%%%%%%%%%%%%%%%%%%%%%%%%%%%%%%%%%%%%%%%%%%%%%%%%%%%%%%%%
%%%%%%%%%%%%%%%%%%%%%%%%%%%%%%%%%%%%%%%%%%%%%%%%%%%%%%%%%%%%%%%%%%%%%%%%%%%%%
%%%%%%%%%%%%%%%%%%%%%%%%%%%%%%%%%%%%%%%%%%%%%%%%%%%%%%%%%%%%%%%%%%%%%%%%%%%%%%
These were performed using $14$ Wannier orbitals for Bi and $16$ Wannier orbitals for Sb and As. Since we consider photoexcited pump pulse photons between $0.5$ eV and $3.0$ eV, we are interested in states within $\sim 2$ eV of the Fermi level, which are well represented by this Wannier interpolation for all three materials. 

The electron-phonon matrix elements are interpolated from a coarse $6 \times 6 \times 6$ grid to finer grids. 
%%%%%%%%%%%%%%%%%%%%%%%%%%%%%%%%%%%%%%%%%%%%%%%%%%%%%%%%%%%%%%%%%%%%%%%%%%%%%%%%%%%%%%%%%
\section{Convergence of $E_g$ force lifetimes}
\begin{figure}[h]
\includegraphics[width=8cm]{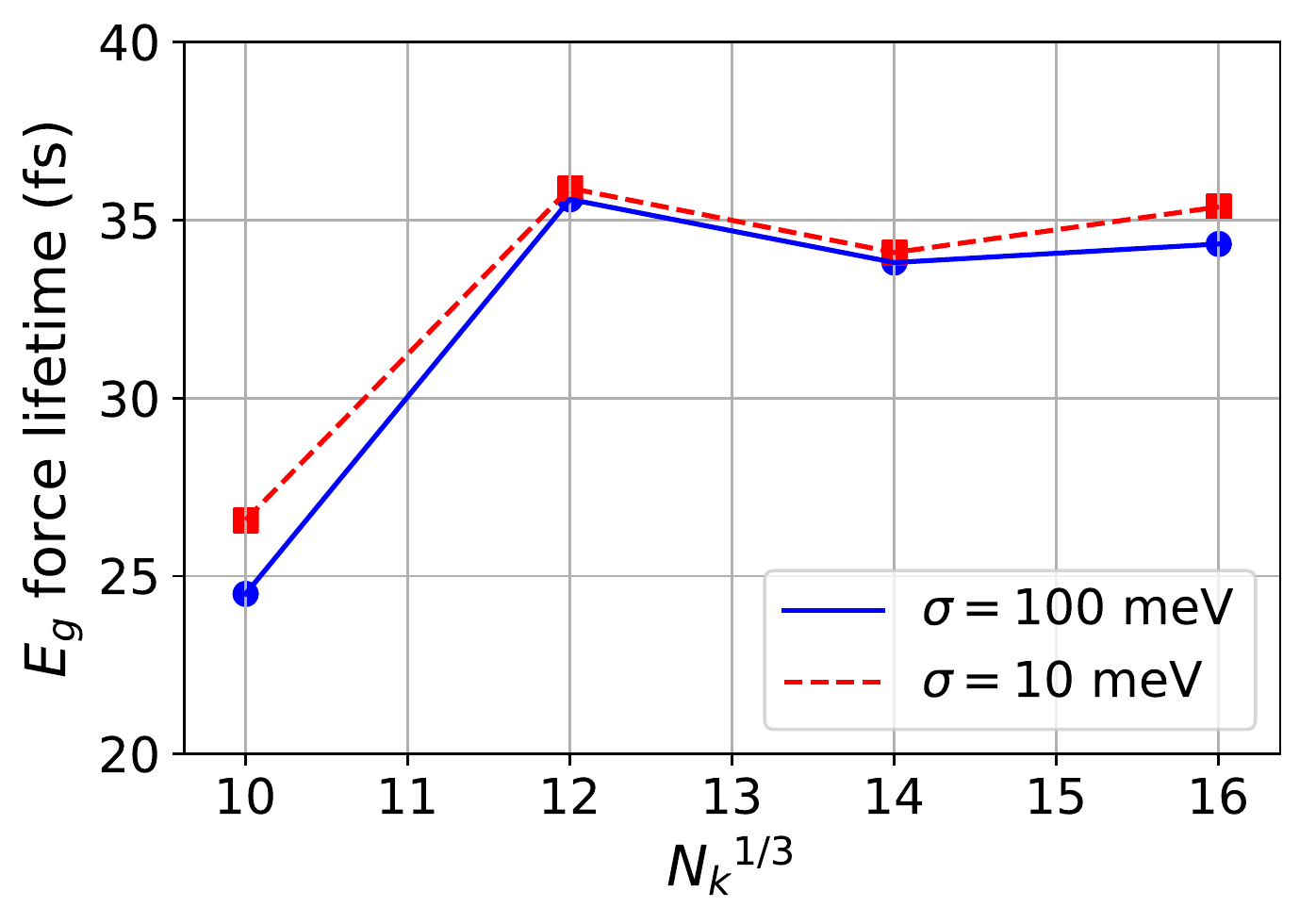}
\caption{Convergence of low-temperature ($0.1$ K) $E_g$ force lifetime in Bi with respect to $N_k$ and $\sigma$ assuming an absorbed photon energy of $1.5$ eV.}
\label{convergence_eg_force}
\end{figure}

The $E_g$ force lifetime has two convergence parameters, the number of $k$(and $q$) points in the uniform Brillouin zone grid ($N_k$) and the Gaussian smearing, $\sigma$, used to compute $\Im{\Sigma_{n \mathbf{k}}}$ (see Eq. $4$ of main text.). As shown in Fig.~$\ref{convergence_eg_force}$, the low-temperature ($0.1$ K) $E_g$ force lifetime at the experimental pump-pulse energy ($1.5$ eV) is insensitive to $\sigma$ and is converged at a Brillouin zone grid of $N_k = 14 \times 14 \times 14$.  The $E_g$ force lifetime converges  at this value of $N_k$ for each of the three materials and at all temperatures considered. 

At pump-pulse energies for which the electrons are being excited to energies at which the electronic density of states is very low, the convergence with respect to grid sampling becomes more demanding. However, for the energy range shown in Fig. $3$ of the main text, going from a $14 \times 14 \times 14$ grid to a $16 \times 16 \times 16$ grid makes at most a difference of $\sim 20 \%$. 
%%%%%%%%%%%%%%%%%%%%%%%%%%%%%%%%%%%%%%%%%%%%%%%%%%%%%%%%%%%%%%%%%%%%%%%%%%%%%%%%%%%%%%%%%%
%%%%%%%%%%%%%%%%%%%%%%%%%%%%%%%%%%%%%%%%%%%%%%%%%%%%%%%%%%%%%%%%%%%%%%%%%%%%%%%%%%%%%%%%%

\section{Antimony $E_g$ force decay rate with additional temperature-independent scattering}

\begin{figure}[h]
\includegraphics[width=8cm]{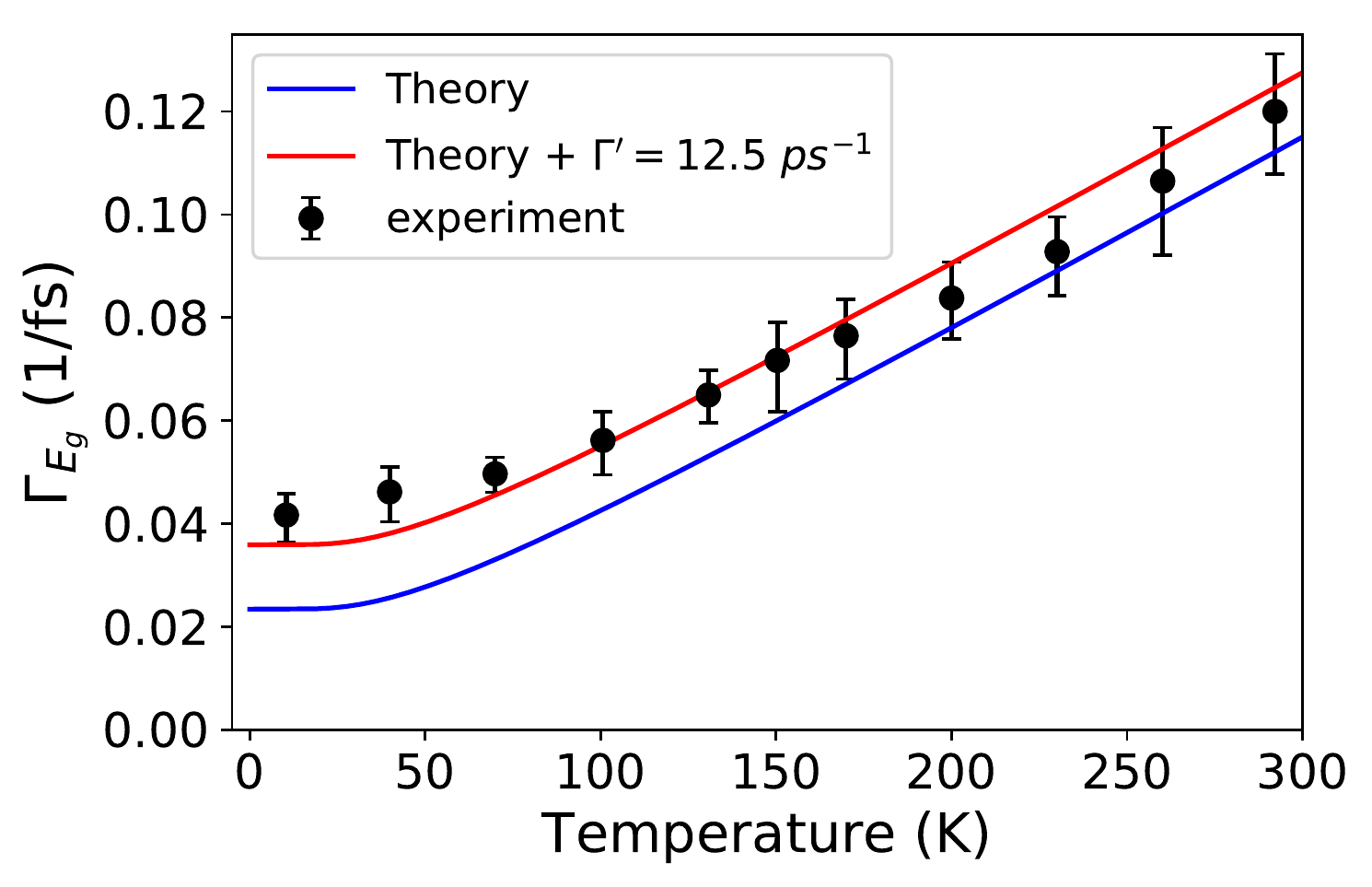}
\caption{Decay rate of $E_g$ force in Sb, including (red line) and excluding(blue line) a correction due to a temperature-independent scattering rate of $\Gamma' \sim 12.5$ $\text{ps}^{-1}$.}
\label{Including_constant_scatt_rate}
\end{figure}

The calculated and experimental values of the Sb $E_g$ force decay rate differ approximately by a temperature-independent scattering rate of $\Gamma' \sim 12.5$ $\text{ps}^{-1}$. Figure.~$\ref{Including_constant_scatt_rate}$ shows the calculated decay rate of the $E_g$ force on Sb, $\Gamma_{E_g}$, the experimental $E_g$ force decay rate and $\Gamma_{E_g} + \Gamma'$. It shows that the discrepancy between the calculated and measured $E_g$ force decay rate in Sb is consistent with a temperature-independent correction due to static imperfections, such as impurities or grain boundaries. 
 
\section{Analysis of experiment including partial decay of $A_{1g}$ force.}
The relaxation time of the $E_g$ mode driving force in bismuth and antimony is experimentally determined by comparing the amplitudes of the $E_g$ and $A_{1g}$ modes obtained from a time-resolved optical pump-optical probe experiment and the corresponding cross sections obtained from cw Raman scattering~\cite{Li2013}. The $A_{1g}$ mode driving force is assumed to remain constant for the duration of the pump pulse ($\sim 70$~fs).  

However, our calculations show a partial decay of the $A_{1g}$ force from its initial value to a non-zero constant value on timescales much shorter than the pump-pulse duration. In order to estimate the impact of this on the experimentally derived $E_g$ force lifetime, we need to understand the amount by which this partial decay modifies the initial amplitude of the $A_{1g}$ mode. 

At times much greater than the pump pulse duration ($t \gg \tau_p$) and assuming that the pump pulse duration is much less than the phonon period ($\Omega \tau_p \ll 1$), we can describe the $A_{1g}$ mode by a harmonic oscillator subject to $2$ driving forces, a step function which decays at a rate $\Gamma$, and one which does not decay. This gives the following equation of motion : 
%%%%%%%%%%%%%%%%%%%%%%%%%%%%%%
\begin{equation}
\ddot{Q} + \Omega^2 Q = \frac{F_0}{\mu}[s + (1 - s)e^{-\Gamma t} ],
\end{equation} 
%%%%%%%%%%%%%%%%%%%%%%%%%%%%%%
where $s \in [0,1]$, $F_0$ is the initial driving force and $\mu$ is the effective mass. Making the substitution $Q \rightarrow \mu Q/F_0$, we arrive at the simpler equation: 
%%%%%%%%%%%%%%%%%%%%%%%%%%%%%%
\begin{equation}
\ddot{Q} + \Omega^2 Q = s + (1 - s)e^{-\Gamma t},
\label{eqn_motion}
\end{equation} 
%%%%%%%%%%%%%%%%%%%%%%%%%%%%%%
which has a general solution of the form:
%%%%%%%%%%%%%%%%%%%%%%%%%%%%%%
\begin{equation}
Q(t) = A \cos(\Omega t + \phi) + \frac{s}{\Omega^2} +  \frac{1-s}{\Gamma^2 + \Omega^2} e^{-\Gamma t}.
\end{equation} 
%%%%%%%%%%%%%%%%%%%%%%%%%%%%%%
The initial conditions are that $Q(0) = 0$ and that $\dot{Q}(0) = 0$, which give us the following:
%%%%%%%%%%%%%%%%%%%%%%%%%%%%%%
\begin{align}
- A \cos(\phi) = \frac{s}{\Omega^2} +  \frac{1-s}{\Gamma^2 + \Omega^2} 
\label{eq:init_cond_1}
\end{align} 
%%%%%%%%%%%%%%%%%%%%%%%%%%%%%%
%%%%%%%%%%%%%%%%%%%%%%%%%%%%%%
\begin{equation}
- A \sin(\phi) =  \frac{1-s}{\Gamma^2 + \Omega^2} \left(\frac{\Gamma}{\Omega}\right).
\label{eq:init_cond_2}
\end{equation} 
%%%%%%%%%%%%%%%%%%%%%%%%%%%%%%
Taking the ratio of these we obtain the phase: 
%%%%%%%%%%%%%%%%%%%%%%%%%%%%%%
\begin{equation}
\tan{\phi} = \frac{\Omega \Gamma (1-s)}{s\Gamma^2 + \Omega^2}.
\end{equation}
%%%%%%%%%%%%%%%%%%%%%%%%%%%%%%
There are two important limits of this expression: when $s=0$, we get $\tan{\phi} = \Gamma/\Omega$, which is the phase of the $E_g$ mode as shown in Ref.~\cite{Li2013}; when $s=1$, we get $\phi = 0$, which is the phase of the $A_{1g}$ mode given by DECP theory. 

Summing the squares of Eq.~$\eqref{eq:init_cond_1}$ and Eq.~$\eqref{eq:init_cond_2}$, we find that the amplitude, $A = \Lambda/\Omega^2$, where $\Lambda$ is defined by: 
%%%%%%%%%%%%%%%%%%%%%%%%%%%%%%
\begin{equation}
\Lambda^2 \equiv \left[ \frac{s^2}{\frac{\Omega^2}{\Gamma^2} + 1} + \frac{1}{\frac{\Gamma^2}{\Omega^2} + 1} \right]
\end{equation}
%%%%%%%%%%%%%%%%%%%%%%%%%%%%%%
This gives us the following equation of motion for the $A_{1g}$ mode: 
%%%%%%%%%%%%%%%%%%%%%%%%%%%%%%
\begin{equation}
Q(t) = \frac{1-s}{\Gamma^2 + \Omega^2} e^{-\Gamma t} + \frac{\Lambda}{\Omega^2} \left[ \frac{s}{\Lambda} - \cos(\Omega t + \phi) \right]
\label{displ_partial_decay}
\end{equation} 
%%%%%%%%%%%%%%%%%%%%%%%%%%%%%%
If we compare this with the equation of motion for the $A_{1g}$ mode driven by a \textbf{time-independent} force: 
%%%%%%%%%%%%%%%%%%%%%%%%%%%%%%
\begin{equation}
Q(t) = \frac{1}{\Omega^2} \left[1 -  \cos(\Omega t + \phi) \right], 
\label{no_partial_decay}
\end{equation} 
%%%%%%%%%%%%%%%%%%%%%%%%%%%%%%
we see that the effect of the force decaying from $F_0 \rightarrow sF_0$ is to reduce the amplitude of the $A_{1g}$ mode by the factor $\Lambda$. 

\section{Effect on derived experimental $E_g$ force lifetime}
%%%%%%%%%%%%%%%%%%%%%%%%%%%%%%%%%%%%%%%%%%%%%%%%
Li et. al. gives the following expression for the $E_g$ force relaxation rate~\cite{Li2013}: 
%%%%%%%%%%%%%%%%%%%%%%%%%%%%%%
\begin{equation}
\Gamma_{E_g} = \Omega_{E_g} \sqrt{ \frac{{g_{PP}}^4}{{g_{RS}}^4} - 1}, 
\label{eq:PRL_Eg_force_lifetime}
\end{equation} 
%%%%%%%%%%%%%%%%%%%%%%%%%%%%%%
where ${g_{PP}}^4 = (A_{A_{1g}} \tilde A_{A_{1g}}/A_{E_g} \tilde A_{E_g})^2$ is the "effective electron-phonon coupling" from the optical pump-optical probe experiment and ${g_{RS}}^4 = (A_{A_{1g}}/A_{E_g})^4$ is the corresponding coupling deduced from cw Raman scattering cross sections which are insensitive to electronic decay of the mode driving forces. The amplitudes are assumed to be of the form~\cite{Li2013}: 
\begin{align}
&A_{E_g} = \frac{F^0_{E_g}}{\mu {\Omega_{E_g}}^2} \\
&\tilde A_{E_g} = \frac{F^0_{E_g}}{\mu {\Omega_{E_g}}^2 \sqrt{1 + \frac{{\Gamma_{E_g}}^2}{{\Omega_{E_g}}^2}}} \\
&A_{A_{1g}} = \frac{F^0_{A_{1g}}}{\mu {\Omega_{A_{1g}}}^2}.
\end{align}
The amplitude $\tilde A_{A_{1g}}$ is assumed to be approximately equal to $A_{A_{1g}}$, which amounts to assuming that the $A_{1g}$ driving force remains constant over the duration of the pump pulse ( $\sim 70$~fs). Since our calculations show a partial decay of the $A_{1g}$ driving force in Bi and Sb, we make the following modification:
%%%%%%%%%%%%%%%%%%%%%%%%%%%%%%
\begin{equation}
\tilde A_{A_{1g}} = \frac{F^0_{A_{1g}} \Lambda}{\mu {\Omega_{A_{1g}}}^2},
\end{equation}
%%%%%%%%%%%%%%%%%%%%%%%%%%%%%%
which implies a change to the derived values of the $E_g$ force lifetime by a factor of:
\begin{equation}
\tau_{E_g} = \frac{ \sqrt{ \frac{{g_{PP}}^4}{{g_{RS}}^4} - 1}}{\sqrt{ \Lambda^2 \frac{{g_{PP}}^4}{{g_{RS}}^4} - 1}}.
\end{equation} 
%%%%%%%%%%%%%%%%%%%%%%%%%%%%%%
Fig.~$\ref{adjust_experiment}$ shows the resulting modifications to the experimentally derived $E_g$ force lifetimes for Bi and Sb: 
\begin{figure}[h]
\includegraphics[width=8cm]{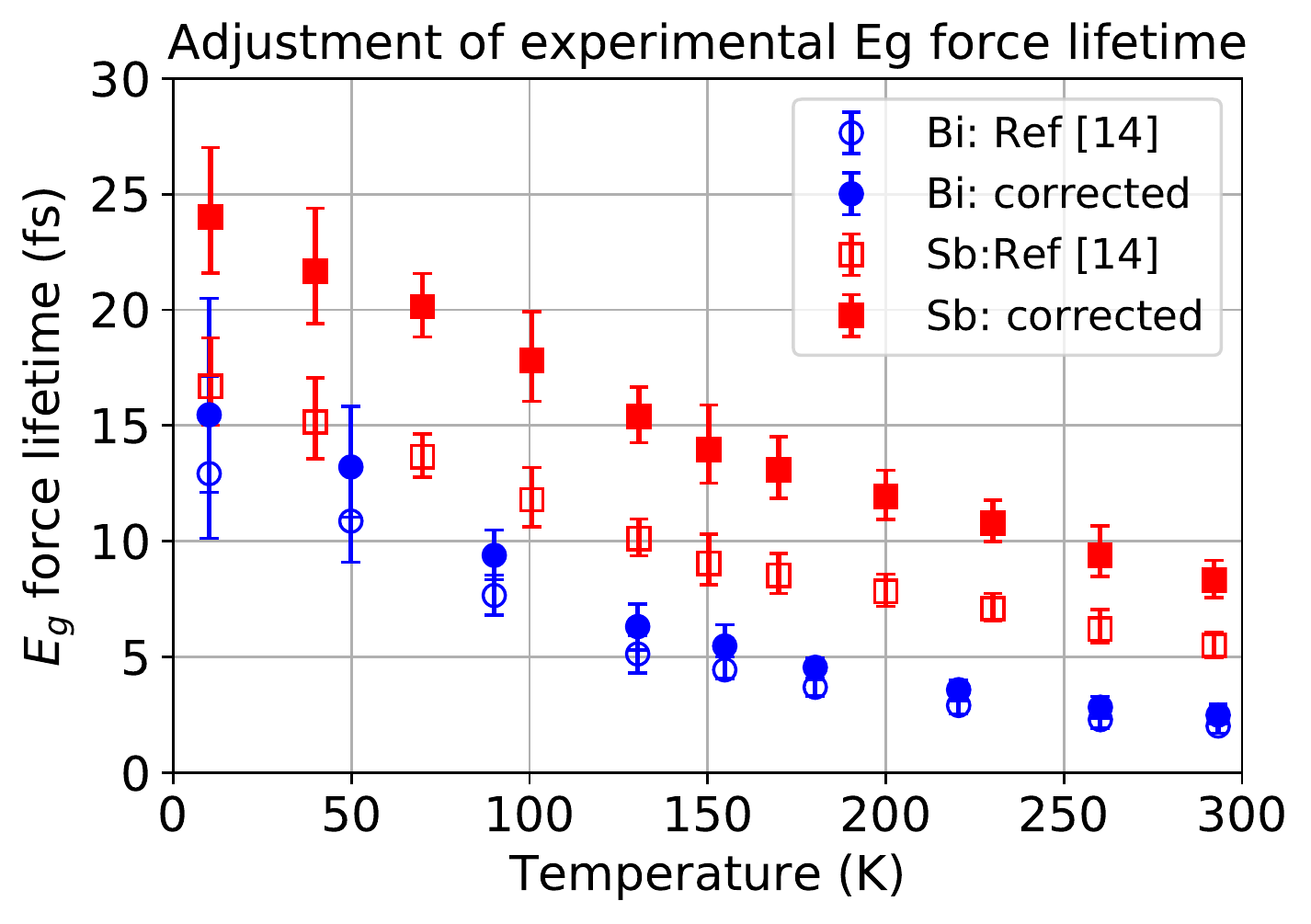}
\caption{Corrections to the experimentally derived $E_g$ force lifetimes reported in Ref \cite{Li2013}. The open points are the experimental values reported in Ref \cite{Li2013}, the solid points are the experimental values after taking into account the calculated partial decay of the $A_{1g}$ driving force.} 
\label{adjust_experiment}
\end{figure}

\section{$E_{g}$ and $A_{1g}$ forces from electron-phonon matrix elements}
%%%%%%%%%%%%%%%%%%%%%%%%%%%%%%%%%%%%%%%%%%%%%%%%%%%%%%
Koopman's theorem relates the DFT total energy, $E_{\text{DFT}}$ to the energy and occupation of orbital $\ket{n \mathbf{k}}$~\cite{martin_2004}:
\begin{equation}
    \frac{\partial E_{\text{DFT}}}{\partial f_{n \mathbf{k}}} = E_{n}(\mathbf{k}) = \mel{ n \mathbf{k}} {\hat{H}}{n \mathbf{k}}.
\end{equation}
To first order in $\Delta f_{n \mathbf{k}}$, the change in the DFT total energy per unit cell is as follows: 
\begin{align*}
    \Delta E_{\text{DFT}/\text{cell}} &\approx \frac{1}{N}  \sum_{n,\mathbf{k}} \Delta f_{n \mathbf{k}} \mel{ n \mathbf{k}} {\hat{H}}{n \mathbf{k}} \\
     &= \frac{1}{N} \sum_{n,\mathbf{k}}  \left(f_{n \mathbf{k}} - f_{n \mathbf{k}}^{(0)}\right) \mel{ n \mathbf{k}} {\hat{H}}{n \mathbf{k}} \numberthis
\end{align*}
where $f_{n \mathbf{k}}^{(0)}$ is the equilibrium electronic occupation of the state $\ket{n \mathbf{k}}$ and N is the number of unit cells in the system. The force $\bm{F_{\alpha}}$ on atom $\alpha$ is then:  
\begin{align*}
    F_{\alpha i} &= - \nabla_{\bm{\tau}_{\alpha}}[\Delta E_{\text{DFT}}/\text{cell}] \\
    &= - \frac{1}{N} \sum_{n,\mathbf{k}} \nabla_{\bm{\tau}_{\alpha}} \left[\Delta f_{n \mathbf{k}} \mel{ n \mathbf{k}} {\hat{H}}{n \mathbf{k}}\right] \\
    &\approx  - \frac{1}{N} \sum_{n,\mathbf{k}} \Delta f_{n \mathbf{k}} \nabla_{\bm{\tau}_{\alpha}} \left[\mel{ n \mathbf{k}} {\hat{H}}{n \mathbf{k}}\right]
\end{align*}
If we assume that the single-particle states $\ket{n \mathbf{k}}$ are eigenstates of $\hat{H}$, then we can apply the Hellman-Feynman theorem~\cite{Feynman1939}. This allows us to express the forces in terms of the diagonal electron-phonon matrix elements and the occupations of the electronic states: 
\begin{equation}
    \boxed{F_{\alpha i} = - \frac{1}{N} \sum_{n,\mathbf{k}} \Delta f_{n \mathbf{k}}  \mel{ n \mathbf{k}} {\nabla_{\bm{\tau}_{\alpha}} \hat{H}}{n \mathbf{k}}}
\end{equation}
\vspace{1mm}

%\bibliography{PRL_Eg_force_decay}{}

\end{document}